\documentclass[aps,amsmath,preprint, prl]{revtex4}
\usepackage[dvips]{graphicx}
\usepackage{braket}
\usepackage{bm}
\usepackage{ulem}
\usepackage[dvips]{color}
\usepackage{setspace}[10]
\usepackage[dvips]{color} 
\def\rnum#1{\expandafter{%
\romannumeral #1}}
\def\Rnum#1{\uppercase\expandafter{%
\romannumeral #1}}

\newcommand{\bol}[1]{\boldsymbol #1}

\begin{document}
\title{Observation of spin current in quantum spin liquid}
\author{Daichi Hirobe}
\email{daichi.kinken@imr.tohoku.ac.jp}
\affiliation{Institute for Materials Research, Tohoku University, Sendai 980-8577, Japan}
\author{Masahiro Sato}
\affiliation{Spin Quantum Rectification Project, ERATO, Japan Science and Technology Agency, Sendai 980-8577, Japan}
\affiliation{The Advanced Science Research Center, Japan Atomic Energy Agency, Tokai 319-1195, Japan}
\author{Takayuki Kawamata}
\affiliation{Department of Applied Physics, Tohoku University, Sendai 980-8579, Japan}
\author{Yuki Shiomi}
\affiliation{Institute for Materials Research, Tohoku University, Sendai 980-8577, Japan}
\affiliation{Spin Quantum Rectification Project, ERATO, Japan Science and Technology Agency, Sendai 980-8577, Japan}
\author{Ken-ichi Uchida}
\affiliation{Institute for Materials Research, Tohoku University, Sendai 980-8577, Japan}
\affiliation{PRESTO, Japan Science and Technology Agency, Saitama 332-0012, Japan}
\author{Ryo Iguchi}
\affiliation{Institute for Materials Research, Tohoku University, Sendai 980-8577, Japan}
\affiliation{Spin Quantum Rectification Project, ERATO, Japan Science and Technology Agency, Sendai 980-8577, Japan}
\author{Sadamichi Maekawa}
\affiliation{Spin Quantum Rectification Project, ERATO, Japan Science and Technology Agency, Sendai 980-8577, Japan}
\affiliation{The Advanced Science Research Center, Japan Atomic Energy Agency, Tokai 319-1195, Japan}
\author{Yoji Koike}
\affiliation{Department of Applied Physics, Tohoku University, Sendai 980-8579, Japan}
\author{Eiji Saitoh}
\email{eizi@imr.tohoku.ac.jp}
\affiliation{Institute for Materials Research, Tohoku University, Sendai 980-8577, Japan}
\affiliation{Spin Quantum Rectification Project, ERATO, Japan Science and Technology Agency, Sendai 980-8577, Japan}
\affiliation{The Advanced Science Research Center, Japan Atomic Energy Agency, Tokai 319-1195, Japan}
\affiliation{WPI Advanced Institute for Materials Research, Tohoku University, Sendai 980-8577, Japan}
\date{\today}
\maketitle
\title{Abstract}
\textbf{Spin liquid is a state of electron spins where quantum fluctuation breaks magnetic ordering with keeping spin correlation\cite{Balents}. It has been one of central topics of magnetism because of its relevance to fascinating phenomena such as high-\mbox{\boldmath $T_\mathbf{\boldsymbol c}$} superconductivity\cite{Anderson, Lee} and topological states\cite{Wen}. In spite of the profound physics, on the other hand, spin liquid itself has been quite difficult to utilize. Typical spin liquid states are realized in one-dimensional spin systems, called quantum spin chains\cite{Giamarchi, Schlappa}. Here we show that a spin liquid in a spin-1/2 quantum chain generates and carries spin current via its long-range spin fluctuation. This is demonstrated by observing an anisotropic negative spin Seebeck effect\cite{Uchida08, Jaworski, Uchida10, Kikkawa, Rezende, Kikkawa15} along the spin chains in $\mathbf{\boldsymbol Sr_2CuO_3}$. The result shows that spin current can flow even in an atomic channel owing to the spin liquid state, which can be used for atomic spin-current wiring.}
 
\title{Introduction}
A flow of electrons’ spin angular momentum is called spin current\cite{Maekawa}. In condensed matter science, transport properties of spin current have attracted considerable interest since the discovery of various spin-current phenomena\cite{Slonczewski, Berger}. In spintronics\cite{Zutic}, on the other hand, it is of critical importance to find materials which can carry spin angular momentum efficiently in integrated microscopic devices.
 
Two types of spin current have experimentally been explored so far. The first one is conduction-electron spin current, which is mediated by an electron motion in metals and semiconductors. Its velocity and propagation length are thus limited by electron diffusion\cite{Bass}. The other type is spin-wave spin current\cite{Kajiwara, Cornelissen}, where spin waves, wave-like propagation of spin motions in magnets, carry spin angular momentum. Its excitation gap is equal to a spin-wave gap, proportional to magnetic anisotropy. Importantly, spin-wave spin current can exist even in insulators in which spin relaxation via conduction electrons is absent, an advantage which may realize fast and long-range spin current transmission, opening a new field of insulator-based spintronics. However, spin-wave spin current in classical magnets may not be suitable for microscopic devices, since handling spin waves becomes difficult when devices are miniaturized toward atomic scale; in ferromagnets, spontaneous magnetization brings about significant stray fields, causing crosstalk. In an antiferromagnetic system, on the other hand, spin ordering patterns should be broken or interfered when a device is in atomic scale; in both cases, spin waves become vulnerable. Therefore, to realize spin-current transport in microscopic devices, spin ordering is expected to vanish with keeping strong interaction among spins.
 
Here, we would like to make a new type of spin current debut: spinon spin current, which may provide a channel for atomic spin transmission to satisfy the requirements. A spinon generally refers to magnetic elementary excitation in quantum spin liquid states\cite{Balents}. When system size of a magnet is reduced to atomic scale, quantum spin fluctuation comes up to the surface and dominates spin properties. The most typical one is found in one-dimensional spin-1/2 chains realized in some oxides, such as an insulator $\mathrm{Sr_2CuO_3}$\cite{Motoyama, Sologubenko, Kawamata}. In $\mathrm{Sr_2CuO_3}$, as shown in Fig. 1a, each $\mathrm{Cu}^{2+}$ ion carries spin-1/2 and is connected each other linearly along the $b$-axis. Because of the one-dimensionality, fluctuation of the spin-1/2 is so strong that it prevents magnetic ordering. As a result, antiferromagnetic interaction embedded in the chain creates a paramagnetic state accompanied by strong spin-singlet correlation, called a quantum spin liquid state. Spin excitation from the spin-liquid ground state has been predicted to be particle-like and to exhibit zero excitation gap: this excitation is called a spinon. This gapless feature is robust against magnetic fields and magnetic anisotropy. Furthermore, theories have predicted that the correlation of spinons is of a markedly long-length scale; even infinite correlation length is predicted in the context of the Tomonaga-Luttinger liquid theories\cite{Giamarchi}. These mean that, in such a system, spin current may propagate in a long distance via spinons along the atomic chain: an ideal feature for atomic spin-current interconnection.
 
To drive spin current, one of the most versatile methods is to use a longitudinal spin Seebeck effect (LSSE)
. LSSE refers to generation of spin current as a result of a temperature gradient applied across a junction between a magnet, typically an insulator magnet, and a metal film, typically Pt. The temperature gradient injects spin current into the metal from the magnet. The injected spin current is converted into electric voltage via the inverse spin Hall effect (ISHE)\cite{Saitoh, Valenzuela, Kimura} in the metal. The voltage is generated perpendicular to the spin polarization and the propagation directions of the spin current. By measuring the voltage generation, the method enables sensitive detection of spin current. The amplitude of the injected spin current is proportional to the non-equilibrium accumulation of spin angular momentum at the interface in the magnet. In the present study, we utilized LSSE to extract spin current from a spin liquid system.
 
Spinon-induced LSSE is characterized by a distinguished feature: theory predicts that the sign of angular momentum due to a  spinon LSSE is opposite to that of the spin-wave LSSE at low temperatures under magnetic fields. The opposite angular momentum is due mainly to the singlet correlation growing with lowering temperature in the spin liquid states in contrast to ferromagnetic correlation growing in classical magnets. Detailed theoretical calculations using a microscopic model reproduce this intuition, which is described in the following. By exploiting these properties of LSSE, we observed spin-current generation and transmission in $\mathrm{Sr_2CuO_3}$.

   Figure 1b is a schematic illustration of the experimental set-up used in the present study. The sample consists of a single crystal of $\mathrm{Sr_2CuO_3}$ and a Pt thin film. The Pt film is used as a spin-current detector based on ISHE, in which spin current is converted into an electromotive force, $\mathbf{E}_\mathrm{SHE}$ (Fig. 1d). The spin chains in the $\mathrm{Sr_2CuO_3}$ are set normal to the Pt film plane (Fig. 1c). A temperature gradient, $\nabla T$, was generated along the spin chains by applying the temperature difference $\it{\Delta}T$ between the top of the Pt film and the bottom of the $\mathrm{Sr_2CuO_3}$ (see also Fig. 1b). Voltage difference, $V$, is measured between the ends of the Pt film with applying an in-plane field, $\mathbf{B}$. 

   First, we measured $\it{\Delta}T$-induced voltage in a Pt film without $\mathrm{Sr_2CuO_3}$. In this simple film, voltage is produced via the normal Nernst effect of Pt alone\cite{Wu}. In Fig. 1e, we show the magnetic field $B$ dependence of the voltage at several temperatures. The voltage $\widetilde{V}=V/\it{\Delta}T$ (the voltage $V$ divided by the temperature difference $\it{\Delta}T$) was found to be proportional to $B$. In Fig. 1f, we show the temperature $T$ dependence of the slope $\widetilde{V}/B$ (i.e. the Nernst coefficient of the Pt). The sign of $\widetilde{V}/B$ is positive through the whole range of $T$, showing that the sign of the normal Nernst effect of Pt is positive in the whole temperature range. 

   The temperature dependence of $\widetilde{V}/B$ for Pt changes dramatically when $\mathrm{Sr_2CuO_3}$ is attached to the Pt. Figure 3a shows the $T$ dependence of $\widetilde{V}/B$ for Pt/$\mathrm{Sr_2CuO_3}$. The sign of $\widetilde{V}/B$ is positive around room temperature, the same sign as the normal Nernst effect in the simple Pt film. With decreasing $T$, surprisingly, the sign of $\widetilde{V}/B$ reverses around 180 K and is negative below the temperature (see also Figs. 2c and 2d). This sign reversal shows that a negative-sign $\widetilde{V}/B$ component appears by attaching an insulator $\mathrm{Sr_2CuO_3}$ and it dominates at low temperatures. Clearly, the negative sign of $V/\it{\Delta}T$ cannot be explained by the normal Nernst effect of Pt, but it is the very feature of the aforementioned spinon LSSE; the sign of the LSSE voltage for Pt/ferro- or ferri-magnets are the same as that of the normal Nernst effect of Pt\cite{Uchida10, Kikkawa}.

   The sign reversal of $\widetilde{V}/B$ was found to be related to spin-current injection from $\mathrm{Sr_2CuO_3}$ as follows. In Fig. 3a, $\widetilde{V}/B$ measured for W/$\mathrm{Sr_2CuO_3}$ is plotted as a function of $T$ (red data points), where W exhibits negative ISHE; the sign of ISHE of W is opposite to that of Pt\cite{Hoffmann}. In W/$\mathrm{Sr_2CuO_3}$, $\widetilde{V}/B$ is always positive and does not exhibit any sign reversal (see also Figs. 3c and 3d), and, remarkably, $V/\it{\Delta}T$ peak with positive sign appears around 20 K (pink arrow in Fig. 3a): the opposite peak sign to that of Pt/$\mathrm{Sr_2CuO_3}$ (sky blue arrow in Fig. 3a). The sign change between W and Pt shows that the low-temperature $V/\it{\Delta}T$ signal is attributed mainly to ISHE due to spin current injected from $\mathrm{Sr_2CuO_3}$.

   In Fig. 3e, $\widetilde{V}/B$ was compared between the $\nabla T\parallel b$-axis and the $\nabla T\perp b$-axis configurations. The $b$-axis is the spin-chain direction of $\mathrm{Sr_2CuO_3}$, and thus $\nabla T\parallel b$-axis ($\nabla T\perp b$-axis) refers to the spin-angular-momentum condition under which the heat current flows parallel (normal) to the spin chains. Clearly, the negative $\widetilde{V}/B$ peak observed in Pt/$\mathrm{Sr_2CuO_3}$ is suppressed when $\nabla T\perp b$-axis: the amplitude of $\widetilde{V}/B$ at 20 K is one order of magnitude less than the $\nabla T\parallel b$-axis configuration (see also Figs. 3g and 3h). The suppression was confirmed also in W/$\mathrm{Sr_2CuO_3}$ (see the inset to Fig. 3e). The result shows that the spin-current injection from $\mathrm{Sr_2CuO_3}$ takes place only when heat current is applied along the spin chain; spin angular momentum flowing along the spin chain of $\mathrm{Sr_2CuO_3}$ dominates the spin-current injection observed in the present study. The tiny negative signal of $\widetilde{V}/B$ for $\nabla T\perp b$-axis (Fig. 3e) might be attributed to an inevitable slight misalignment of $\nabla T$ direction from the $b$-axis ($\leq6^\circ$). We also note that the thermal conductivity of $\mathrm{Sr_2CuO_3}$ is almost isotropic\cite{Kawamata} and, therefore, the voltage suppression cannot be attributed to a reduction in the magnitude of $\nabla T$. The negative and anisotropic LSSE is evidence that spin current is generated and conveyed by spinons through the spin chains of the $\mathrm{Sr_2CuO_3}$. 

   The spin transport along the spin chain was confirmed also by disappearance of $\widetilde{V}/B$ by partially breaking the spin chains. The negative $\widetilde{V}/B$ signal in Pt/$\mathrm{Sr_2CuO_3}$ disappeared when the average spin-chain length was reduced from $\sim10^{-6}$ m to $\sim10^{-7}$ m by decreasing the purity of the starting compounds of $\mathrm{Sr_2CuO_3}$ (99.999\% $\rightarrow$ 99.9\%)\cite{Hlubek}.

   Finally, we turn to theoretical formulation of the spinon LSSE in the present system. We calculated a spin current, $I_\mathrm{s}$, injected across the metal/$\mathrm{Sr_2CuO_3}$ interface by combining a Tomonaga-Luttinger liquid theory\cite{Giamarchi} with a microscopic theory for LSSE\cite{Adachi11} (see Supplementary Sections SA and SB). Figure 4 shows a calculated field dependence of $I_\mathrm{s}$ injected from $\mathrm{Sr_2CuO_3}$. We also show a result for the ferromagnetic LSSE obtained by calculating $I_\mathrm{s}$ for a three-dimensional ferromagnet (e.g. $\mathrm{Y_3Fe_5O_{12}}$) (see Supplementary Section SC). $I_\mathrm{s}$ injected from $\mathrm{Sr_2CuO_3}$ is proportional to the external magnetic field and, importantly, the sign of the spinon $I_\mathrm{s}$ is opposite to that of the spin-wave $I_\mathrm{s}$: the key feature observed experimentally. In addition, the magnitude of the calculated $I_\mathrm{s}$ at 2 T (spinon $I_\mathrm{s}$ $\sim10^{-4}\times$spin-wave $I_\mathrm{s}$) is fairly consistent with the experimental values\cite{Kikkawa15}.

   Recently, optically induced crystallization of amorphous Sr-Cu-O was developed\cite{Takahashi}. In the crystallization, spin-chain directions were found to align along the light-scanning direction, an advantage in application to tailor-made spin wiring. We anticipate that the present discovery of spin-current transmission along a quantum spin chain will also lead to advances of spin integrated circuit technology.

\section*{Methods}	
\subsection*{Sample preparation}
\noindent The single crystalline $\mathrm{Sr_2CuO_3}$ was grown from primary compounds $\mathrm{SrCO_3}$ and $\mathrm{CuO}$ with 99.999\% by a traveling-solvent floating-zone method\cite{Kawamata}. The single crystalline $\mathrm{Sr_2CuO_3}$ was cut into a cuboid 5 mm long, 1 mm wide, and 1 mm thick. The surface of the $\mathrm{Sr_2CuO_3}$ was polished mechanically in a glove box filled with a $\mathrm{N_2}$ gas. We found that exposure of the sample to air causes deterioration of the sample, since the surface of $\mathrm{Sr_2CuO_3}$ reacts rapidly with moisture\cite{Hlubek}. The 7-nm-thick Pt film was then sputtered on the polished surface ($5\times1$ $\mathrm{mm}^2$) of the $\mathrm{Sr_2CuO_3}$ in an Ar atmosphere. 
\subsection*{Voltage measurement}
\noindent Voltage data were taken in a Physical Properties Measurement System (Quantum Design, Inc.). The Pt (W)/$\mathrm{Sr_2CuO_3}$ sample was sandwitched by sapphire plates and the bottom of the sample was thermally anchored at the system temperature $T$. The temperature gradient $\nabla T$ was generated between the sapphire plates using a chip resistor (100 $\Omega$) on the sapphire plate attached to the metal film. The temperature difference $\it{\Delta}T$ between the sapphire plates was set to be ${\it{\Delta}}T/T<0.1$ at each system temperature $T$. Two electrodes were attached to both the ends of the metal film to measure voltage. An exter	nal magnetic field was applied in the in-plane direction which is normal to the direction of $\nabla T$ as well as the direction across the two electrodes. 
\section*{Acknowledgements}
\noindent The authors thank N. Yokoi and K. Satoh for valuable discussions. This work was supported by 
ERATO-JST `Spin Quantum Rectification Project', Japan, 
Grant-in-Aid for Scientific Research on Innovative Area ``Nano Spin Conversion Science'' (No. 26103005), 
PRESTO ``Phase Interfaces for Highly Efficient Energy Utilization'' from JST, Japan, 
Grant-in-Aid for Challenging Exploratory Research (No. 26610091), 
Grant-in-Aid for Challenging Exploratory Research (No. 26600067), and
Grant-in-Aid for Scientific Research (A) (No. 15H02012) from MEXT, Japan. 
D. H. was supported by Yoshida Scholarship Foundation through the Doctor 21 program.
\section*{Author contributions}
\noindent D. H. and E. S. designed the experiments; T. K. grew single crystals used in the study; D. H. collected and analysed the data; K. U. and R. I. supported the experiments; M. S. developed the theoretical explanations; S. M., Y. K. and E. S. supervised the study; D. H., M. S., Y. S. and E. S wrote the manuscript. All authors discussed the results and commented on the manuscript.
\section*{Competing financial interests}
\noindent The authors declare no competing financial interests.

\newpage
\noindent\textbf{Figure 1 $\mathbf{|}$ Quantum spin chain, experimental set-up, and thermoelectric response of Pt film without $\mathbf{\boldsymbol Sr_2CuO_3}$.}
\textbf{a}, Quantum spin chains in $\mathrm{Sr_2CuO_3}$. Quantum spin chains along the $b$-axis consist of $\mathrm{Cu^{2+}}$ ions sharing $\mathrm{O^{2-}}$ ions. \textbf{b}, A schematic illustration of the experimental set-up. The sample consists of single crystalline $\mathrm{Sr_2CuO_3}$ and a Pt film. A temperature gradient, $\nabla T$, is generated along the spin chains ($b$-axis) by applying a temperature difference, ${\it\Delta} T$. $\mathbf{B}$ denotes an external magnetic field and $T$ the system temperature. \textbf{c}, Configuration of quantum spin chains in the experimental set-up. A Cu-O chain is set along the temperature gradient $\nabla T$. \textbf{d}, A schematic illustration of the inverse spin Hall effect. An electric field, $\mathbf{E}_\mathrm{SHE}$, arises in the direction of $\mathbf{J}_{\mathrm{s}}\times\mathbf{\boldsymbol\sigma}$ in Pt. Here $\mathbf{J}_{\mathrm{s}}$ and $\mathbf{\boldsymbol\sigma}$ are the spatial direction and the spin-polarisation direction of spin current, respectively. \textbf{e}, Field ($B$) dependence of $\widetilde{V}=V/{\it\Delta} T$, voltage $V$ divided by the temperature difference ${\it\Delta} T$, in a Pt film put on single crystalline MgO. The dashed lines are fits to the data points. \textbf{f}, Temperature dependence of $\widetilde{V}/B$, a voltage slope ($V/B$) divided by the temperature difference ${\it\Delta}T$, in Pt/MgO. The error bars represent the 68\% confidence level ($\pm$s.d.).

\noindent\textbf{Figure 2 $\mathbf{|}$ Observation of negative spin Seebeck effect in Pt/$\mathbf{\boldsymbol Sr_2CuO_3}$.}
\textbf{a}, Temperature ($T$) dependence of voltage measured in Pt/$\mathrm{Sr_2CuO_3}$. $\widetilde{V}/B$ is the voltage slope $V/B$ divided by the temperature difference ${\it\Delta}T$. The data for Pt/MgO are also presented (see also Fig. 1f). \textbf{b}, Experimental set-ups. In Pt/MgO (upper panel), the electric field $\mathbf{E}_\mathrm{NNE}$ arises via the normal Nernst effect in Pt. In Pt/$\mathrm{Sr_2CuO_3}$ (lower panel), the electric field $\mathbf{E}_\mathrm{SHE}$ also arises via the inverse spin Hall effect in Pt. $\mathbf{B}$ and $\nabla T$ denote the magnetic field and the temperature gradient, respectively. \textbf{c}, \textbf{d}, Field ($B$) dependence of voltage ($V$) measured in the Pt/$\mathrm{Sr_2CuO_3}$ at various temperatures. $\widetilde{V}$ refers to $V$ divided by the temperature difference ${\it\Delta} T$. The lines are fits to the data points. \textbf{e}, \textbf{f}, Dependence of the voltage $V$ on the temperature difference ${\it{\Delta}} T$ at 260 K (\textbf{e}) and at 15 K (\textbf{f}) at 1 T. In both cases, slopes are proportional to ${\it{\Delta}} T$. The ${\it{\Delta}} T$-linear dependence of the positive slope is explained by the normal Nernst effect in Pt. 

\noindent\textbf{Figure 3 $\mathbf{|}$ Spin transport through quantum spin chains.}
\textbf{a}, Temperature ($T$) dependence of voltage measured in W/$\mathrm{Sr_2CuO_3}$. $\widetilde{V}/B$ is the voltage slope $V/B$ divided by the temperature difference ${\it\Delta}T$. The data for Pt/$\mathrm{Sr_2CuO_3}$ are also presented for comparison. \textbf{b}, Experimental set-ups for W/$\mathrm{Sr_2CuO_3}$ (upper panel) and Pt/$\mathrm{Sr_2CuO_3}$ (lower panel). The directions of $\mathbf{E}_\mathrm{SHE}$ are opposite between W and Pt as highlighted by red and blue arrows. $\mathbf{B}$ and $\nabla T$ refer to the magnetic field and the temperature gradient, respectively. \textbf{c}, \textbf{d}, Field ($B$) dependence of voltage measured in W/$\mathrm{Sr_2CuO_3}$ at various temperatures. $\widetilde{V}$ denotes $V$ divided by the temperature difference ${\it\Delta} T$. The lines are fits to the data points. \textbf{e}, Temperature ($T$) dependence of voltage under the temperature gradient $\nabla T$ perpendicular to the spin chains ($a$-axis, green). The result for $\nabla T$ along the spin chains ($b$-axis, blue) is shown for comparison. The inset shows the data for W/$\mathrm{Sr_2CuO_3}$. \textbf{f}, Experimental set-ups for measuring the chain-direction dependence of voltage. The bottom panel shows the directions of $\nabla T$ with respect to the Cu-O chains. \textbf{g}, \textbf{h}, Field ($B$) dependence of voltage ($\widetilde{V}$) measured under $\nabla T$ along the $a$-axis (\textbf{g}) and along the $b$-axis (\textbf{h}) in the Pt/$\mathrm{Sr_2CuO_3}$. The lines are fits to the data points.	

\noindent\textbf{Figure 4 $\mathbf{|}$ Theoretical calculations for spinon and ferromagnetic spin Seebeck effects.} 
External magnetic field ($B$) dependence of spin current ($I_\mathrm{s}$) generated via spinon and ferromagnetic spin Seebeck effects. For the spinon (ferromagnetic) LSSE calculation, exchange coupling, $J$, was set at $-2000$ K ($+50$ K), which is a typical value of $\mathrm{Sr_2CuO_3}$ ($\mathrm{Y_3Fe_5O_{12}}$). A sample temperature, $T$, was set at  20 K. The amplitude of the ferromagnetic $I_\mathrm{s}$ is multiplied by $10^{-4}$.

\newpage
\begin{figure}[htbp]
  \begin{center}
  \includegraphics[width=15.0cm]{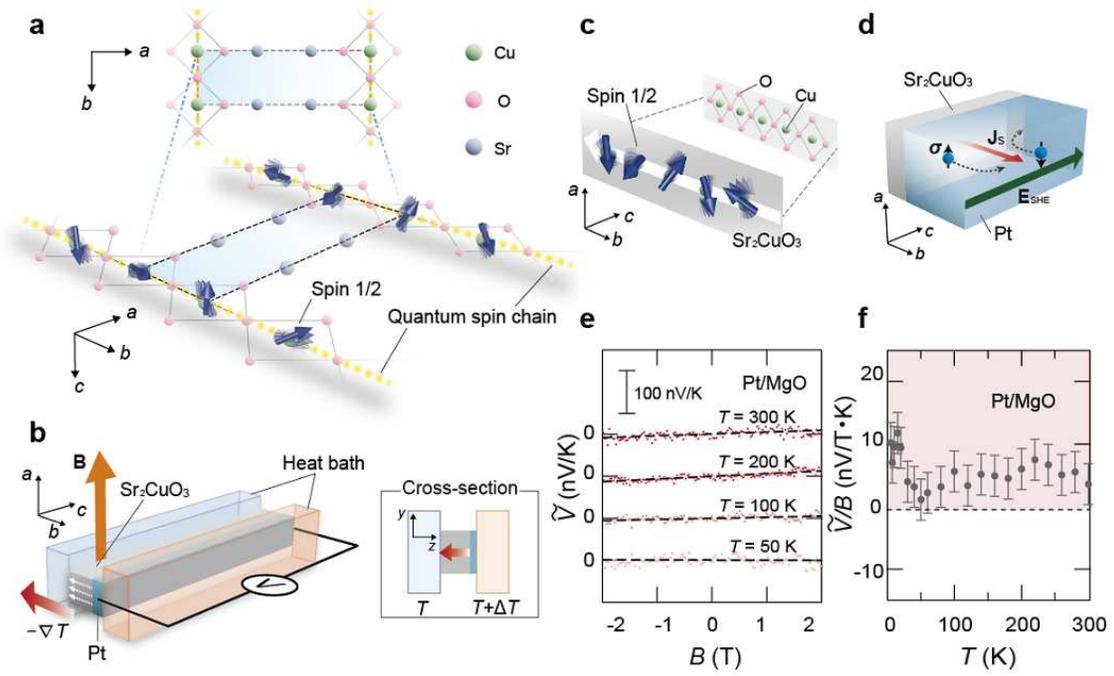}
  \caption{\textbf{Quantum spin chain, experimental set-up, and thermoelectric response of Pt film.}}
  \end{center}
\end{figure}

\newpage{}
\begin{figure}[htbp]
  \begin{center}
  \includegraphics[width=12.0cm]{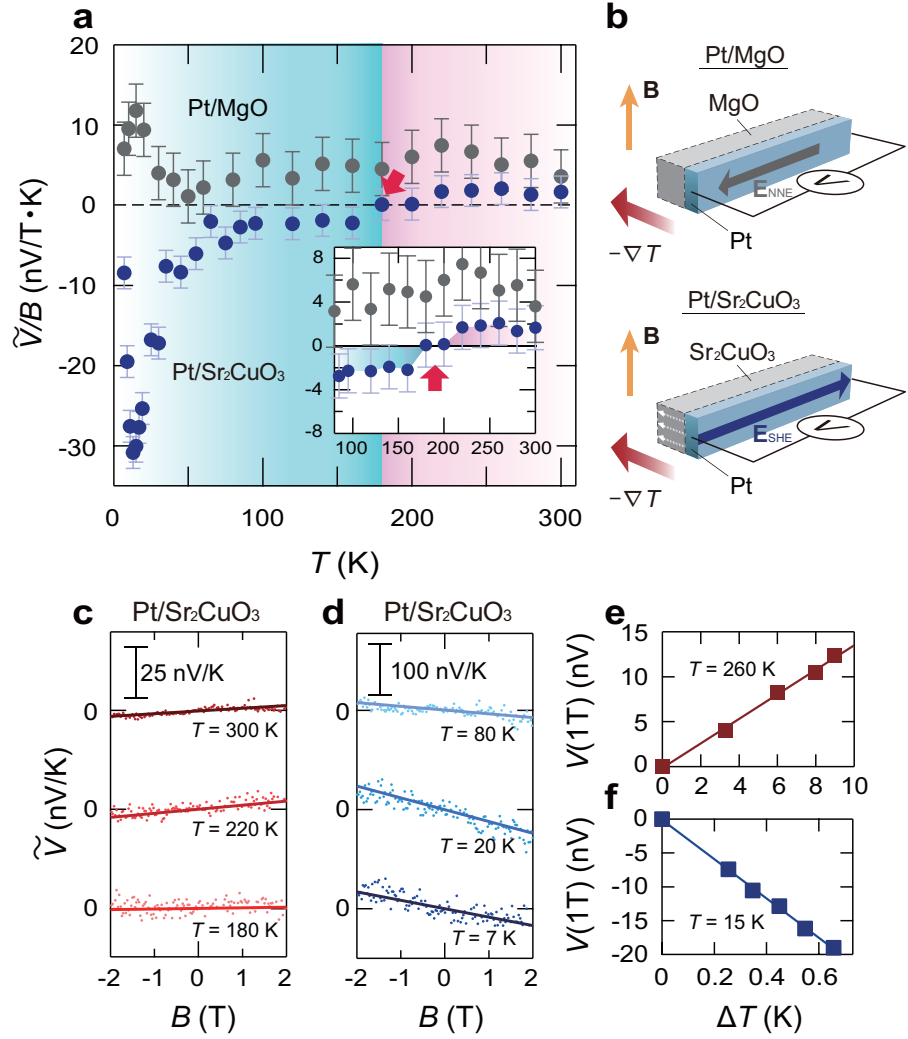}
  \caption{\textbf{Observation of negative spin Seebeck effect in Pt/$\mathbf{\boldsymbol Sr_2CuO_3}$.}}
  \end{center}
\end{figure}

\newpage
\begin{figure}[htbp]
  \begin{center}		
  \includegraphics[width=15.0cm]{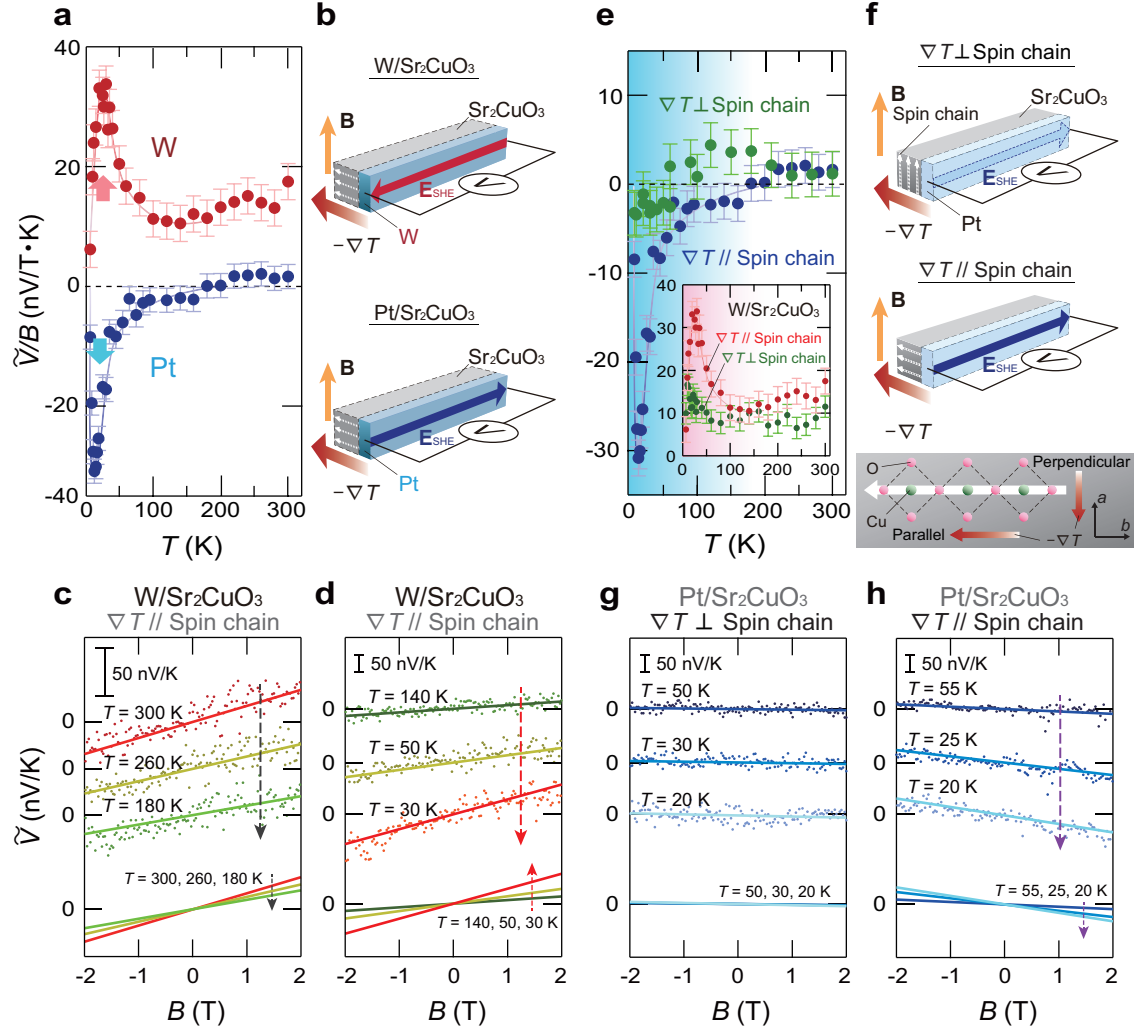}		
  \caption{\textbf{Spin transport along quantum spin chains.}}
  \end{center}
\end{figure}

\newpage
\begin{figure}[htbp]
  \begin{center}		
  \includegraphics[width=8.0cm]{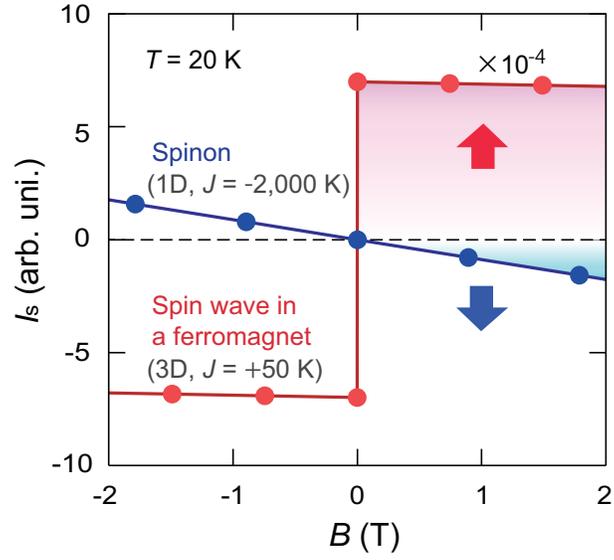}		
  \caption{\textbf{Theoretical calculations for spinon and ferromagnetic spin Seebeck effects.}}
  \end{center}
\end{figure}

\clearpage
\begin{center}
\section*{\large Supplementary Information for \\Observation of spin current in quantum spin liquid}
\end{center}
Daichi Hirobe, Masahiro Sato, Takayuki Kawamata, Yuki Shiomi, Ken-ichi Uchida, Ryo Iguchi, Yoji Koike, Sadamichi Maekawa \& Eiji Saitoh
\\
\subsection*{SA. Antiferromagnetic spin-1/2 chain}
Here, we shortly review the low-energy properties of 
one-dimensional (1D) antiferromagnetic (AF) spin-1/2 chains. 
A typical Hamiltonian of the AF spin-1/2 chain 
is written as 
\begin{eqnarray}
{\cal H}&=& 
-J \sum_j ({S}_{j}^x{S}_{j+1}^x+{S}_{j}^y{S}_{j+1}^y
+\Delta {S}_{j}^z{S}_{j+1}^z)
-B\sum_jS^z_j,
\label{eq:model}
\end{eqnarray}
where ${\bol S}_{j}$ is the spin-1/2 operator 
on the $j$-th site (we set $\hbar=1$), 
$J<0$ is the AF exchange coupling, $\Delta$ is the Ising-type 
anisotropy, and $B=g\mu_B H$ is the magnitude of the external magnetic field 
($g$, $\mu_B$, and $H$ are, respectively, the $g$ factor, the Bohr magneton, 
and the magnetic field). Note that ${\bol S}_j$ is defined as ${\bol S}_j=-{\bol S}_j^\mathrm{e}$ 
(${\bol S}_j^\mathrm{e}$: electron-spin operator) in Eq. (\ref{eq:model}) in accordance with
a standard notation. The point $\Delta=1$ corresponds to the realistic SU(2)-symmetric 
Heisenberg model. Magnetic properties of various quasi-one-dimensional cuprates 
including Sr$_2$CuO$_3$ can be captured by the above model~(\ref{eq:model}). 
The low-energy physics of the spin chain is known to be well 
described by the Tomonaga-Luttinger liquid (TL-liquid) theory~[5, 31, 32] 
with gapless spinon excitations, 
and the gapless spin liquid phase of the Heisenberg model ($\Delta=1$) 
is fairly stable against the magnetic field $B$ and the easy-plane anisotropy 
$|\Delta|<1$. For example, the spin liquid phase survives in the regime 
from zero field to the saturation field $B_c=2J$~[5]. 

\subsection*{SB. Spin Seebeck effect of AF spin-1/2 chain}
\begin{figure}[tth]
\begin{center}
\includegraphics[width=10cm]{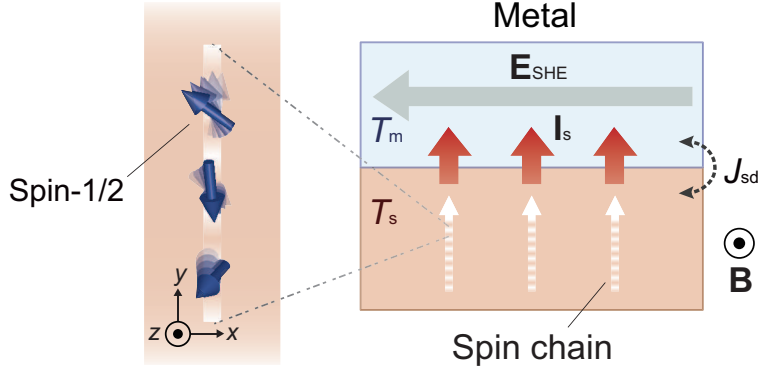}
\end{center}
\caption{Model for spin Seebeck effect in one-dimensional spin chains. 
$T_{\rm m}$ denotes the effective temperature of electrons in the metal; 
$T_{\rm s}$ that of spinons in the one-dimensional spin chains. 
The spin current $\mathbf{I}_{\rm s}$ is injected into the metal via 
the interface exchange interaction with the magnitude $J_{\rm sd}$ 
under the magnetic field $\mathbf{B}$. $\mathbf{I}_{\rm s}$ is converted
into the electromotive force $\mathbf{E}_{\rm SHE}$ via the inverse spin Hall effect.}
\label{fig:Model}
\end{figure}

In this section, we explain the theory part of the longitudinal spin Seebeck 
effect (LSSE) of spin-1/2 AF chains~(\ref{eq:model}). 
We consider the model for the experimental set-up (Fig. 1b), 
as shown in Fig.~\ref{fig:Model}, 
where temperatures of the spin chain (Sr$_2$CuO$_3$) and metal (Pt) are 
respectively set to be $T_{\rm s}$ and $T_{\rm m}$, and 
the exchange coupling $J_{\rm sd}$ exists at the interface. 
For this set-up, the microscopic theory~[28, 33] shows that 
the spin current $I_s$ injected into the metal through the interface 
is given by
\begin{eqnarray}
I_s= -\frac{2N_{\rm int}J_{\rm sd}^2}{\sqrt{2}\pi} 
\int^{\infty}_{-\infty} d\omega 
\,\,{\rm Im}\chi^{-+}(\omega) \,\,{\rm Im}X^{-+}(\omega)\,\,
\big[n(T_{\rm m})-n(T_{\rm s})\big]
\label{eq:spincurrent}
\end{eqnarray}
where $\chi^{-+}$ and $X^{-+}$ are, respectively, 
the local-spin dynamical susceptibilities of the metal and the spin chain, 
$n(T)=1/(e^{\omega/T}-1)$ is the Bose distribution function ($\omega$: frequency), 
and $N_{\rm int}$ is the total number of sites at the interface. 
Indices $-+$ denote 
the transverse spins $S^{\pm}=S^x\pm i S^y$,
and the explicit form of $X^{-+}$ is given by
\begin{eqnarray}
&&X^{-+}(\omega)=\frac{1}{N}\sum_k X^{-+}(\omega,k),
\nonumber\\
&&X^{-+}(\omega,k)=-\int_0^\beta d\tau e^{i\omega_n\tau}
\sum_j e^{-ik j}\langle T_\tau S_j^-(\tau)S_0^+(0)\rangle
\Big|_{i\omega_n\to\omega+i0},
\label{eq:susceptibility}
\end{eqnarray}
where $k$ is the wave number, $N$ is the total number of sites, 
$\tau$ is imaginary time, $\beta=1/T$, $\omega_n=2\pi n/\beta$ ($n$: integer), 
and $T_\tau$ stands for imaginary-time ordered product. 
The susceptibility of metal $\chi^{-+}$ is also defined by 
replacing $j$, $k$ and $S_j^{\pm}$ with 
conduction electron's coordinate $\bol r$, wave vector $\bol k$ and spin 
${s}_{\bol r}^{\pm}$, respectively, in Eq.~(\ref{eq:susceptibility}). 
This interface spin current is converted into an electric current in the metal
via the inverse spin Hall effect~[22, 23], 
and hence the LSSE voltage 
observed in the present study is proportional to the spin current $I_s$. 
Equation~(\ref{eq:spincurrent}) shows that $I_s$ 
vanishes when the temperature difference $\Delta T=T_{\rm s}-T_{\rm m}$ 
disappears. We emphasize that this spin-current 
formula is generally applicable irrespective of any magnetic state 
such as ferromagnetic, antiferromagnetic and spin liquid states.

Let us simplify the spin-current formula~(\ref{eq:spincurrent}). 
The susceptibility of the metal 
may be approximated by a spin-diffusion type function 
${\rm Im}\chi^{-+}(\omega)= \chi_0\omega\tau_s/(1+\tau_s^2\omega^2)$, 
where 
$\chi_0$ is the static susceptibility of the metal, and 
the spin relaxation time $\tau_s$ is almost unchanged 
with changing $T_{\rm m}$ and $B$~[28]. 
Since ${\rm Im}\chi^{-+}$ and the $T$-dependent factor 
$n(T_{\rm m})-n(T_{\rm s})$ are both odd with respect to $\omega$, 
the formula (\ref{eq:spincurrent}) shows that the necessary and 
sufficient condition for generating a finite spin current 
is to make ${\rm Im}X^{-+}$ deviate from the $\omega$-odd function. 
When $\Delta T=T_{\rm s}-T_{\rm m}$ is sufficiently small, 
$n(T_{\rm m})-n(T_{\rm s})$ can be approximated by 
$-\omega \Delta T/(4T^2 \sinh^2(2\omega/T))$, where 
$T=(T_{\rm s}-T_{\rm m})/2$. 
Using these relations of the susceptibility $\chi^{-+}$ and the $T$ factor, 
we can simplify the formula~(\ref{eq:spincurrent}).
The normalized spin current $\tilde I_s$ defined by 
$I_s=-\frac{N_{\rm int}J_{\rm sd}^2}{2\sqrt{2}\pi}\tilde I_s\Delta T$ is given by  
\begin{eqnarray}
\tilde I_s= \frac{1}{T^2}
\int^{\infty}_{-\infty} d\omega 
\,\,{\rm Im}X^{-+}(\omega)\frac{\omega^2}{1+\tau_s^2\omega^2}
\frac{1}{\sinh^2(2\omega/T)}. 
\label{eq:spincurrent2}
\end{eqnarray}

The remaining task is to compute ${\rm Im}X^{-+}$ in 
Eq.~(\ref{eq:spincurrent2}). 
The TL-liquid theory including bosonization~[5, 31, 32] 
and conformal field theory~\cite{Francesco} 
provides a powerful way of calculating dynamical correlation functions of 
TL-liquid phases, and the results are reliable 
in the low-energy and low-temperature regime. The most dominant region of 
the dynamical susceptibility $X^{-+}(\omega,k)$ is located around 
$k=\pi$. According to the TL-liquid theory (see, e.g., Ref.~[5]), 
$X^{-+}$ around $k=\pi$ is given by
\begin{eqnarray}
X^{-+}_{\rm TL}(\omega,\pi+\delta k) &=&
 -B_0^2 \,\,\frac{a_0}{v}\,\,\sin\Big(\frac{\pi}{2K}\Big)
\Big(\frac{2\pi a_0}{\beta v}\Big)^{1/K-2}
B\Big(-i\frac{\beta(\omega-v \delta k)}{4\pi}+\frac{1}{4K},1-\frac{1}{2K}\Big)\nonumber\\
&&\times
B\Big(-i\frac{\beta(\omega+v \delta k)}{4\pi}+\frac{1}{4K},1-\frac{1}{2K}\Big),
\label{eq:susc_TLliquid}
\end{eqnarray}
where $B(x,y)=\Gamma(x+y)/(\Gamma(x)\Gamma(y))$ is the Beta function, 
$v$ is the spinon group velocity, $B_0$ is a non-universal constant, 
and $K>0$ is the TL-liquid parameter. Microscopic information about 
the original spin chain is included in these parameters $v$, $B_0$ and $K$. 
To be specific, these three parameters are functions of $J$, $\Delta$ and $B$. 
At the SU(2)-symmetric Heisenberg point $\Delta=1$, 
the value of $K$ monotonically increases from unity to 2 
as the magnetic field $B$ changes from 0 to the saturation value $2J$. 
Accurate values of $v$, $B_0$ and $K$ were 
determined in Refs.~[5, 35--37]. 
Using them, we obtain the $\omega$-, $k$-, and $T$-dependences of 
the susceptibility $X^{-+}$ from Eq.~(\ref{eq:susc_TLliquid}). 
The large amplitude of $X^{-+}_{\rm TL}$ appears around the linearized 
spinon dispersion curve $\omega=\pm v(k-\pi)=\pm v\delta k$, and 
the spectrum is continuously distributed in $(\omega, k)$ space. 
Equation~(\ref{eq:susc_TLliquid}) shows that 
${\rm Im}X_{\rm TL}^{-+}(\omega,k)$ takes negative (positive) values 
in the positive-$\omega$ (negative-$\omega$) region. 

At $B=0$, it is shown that 
${\rm Im}X^{-+}_{\rm TL}=N^{-1}\sum_{\delta k} {\rm Im}X_{\rm TL}^{-+}
(\omega,\pi+\delta k)=(2\pi)^{-1}\int_0^{\Lambda} dp 
\,\,{\rm Im}X_{\rm TL}^{-+}(\omega,\pi+p)$ 
is an $\omega$-odd function ($\Lambda$: a proper high-energy cut off). 
We therefore obtain zero spin current $\tilde I_s=0$ at $B=0$, and this is 
consistent with the experimental result. The result of 
$\tilde I_s=0$ can be also proven by using time-reversal or 
spin-rotational symmetry. However, 
the formula (\ref{eq:susc_TLliquid}) also shows that even for a finite field 
$B\neq 0$, ${\rm Im}X^{-+}_{\rm TL}$ is odd;
zero spin current at $B\neq 0$ does not agree with the present experimental
results. This suggests that the usual TL-liquid theory is not 
sufficient to explain the LSSE of quantum spin chains. This situation contrasts 
with the fact that the TL-liquid theory has successfully explained other 
dynamical phenomena of 1D magnets such as 
electron spin resonance~\cite{Oshikawa}, 
nuclear magnetic resonance~\cite{Takigawa,Thurber,Sato}, and 
neutron scattering spectra~\cite{Endoh,Tennant,Tennant2}.

In addition to the TL-liquid theory, 
other powerful theoretical techniques have been developed for 1D quantum many-body systems. 
The Bethe ansatz~\cite{Takahashi, Caux, Kohno}, 
one of such techniques, is applicable to the AF 
spin-1/2 chain and it can exactly compute the 
dynamical correlation functions if we restrict ourselves to the $T=0$ case.
It shows that the {\it curved} spinon dispersion 
$\omega=\epsilon(\delta k)$ gives the lower bound of the spectrum 
${\rm Im}X^{-+}(\omega,\pi+\delta k)$ around $k=\pi$ 
at $T=0$. 
This is also supported by numerical calculations~\cite{Shiba,Muller}. 
On the other hand, in the formula~(\ref{eq:susc_TLliquid}), 
(as we mentioned) the spinon dispersion is approximated 
by the linearized dispersion $\omega=\pm v \delta k$. 
Due to this linear dispersion, both the positive- and negative-$\omega$ weights of 
Eq.~(\ref{eq:susc_TLliquid}) cancel out exactly. 
Therefore, a reasonable improvement of Eq.~(\ref{eq:susc_TLliquid}) is 
to replace $\omega\pm v\delta k$ with $\omega -\epsilon(\pm \delta k)$. 
In fact, the recently developed nonlinear TL-liquid theory~\cite{Imambekov} 
also justifies the substitution of the curved spinon dispersion to 
Eq.~(\ref{eq:susc_TLliquid}). 
For $B>0$, the dispersion curve $\omega=\epsilon(\delta k)$ 
becomes flatter in the positive-$\omega$ region than in 
the negative-$\omega$ region. As a result, the contribution from the 
positive-$\omega$ region is dominant in $I_s$, which means that down-spin 
spinons are the main carriers of the spin current. 
We note that (as we will show below) 
magnons carry up-spins in the LSSE of 3D ordered ferromagnets, 
differently from the case of spinons. 
After the substitution of $\omega=\epsilon(\delta k)$ to 
Eq.~(\ref{eq:susc_TLliquid}), we finally arrive at a finite, {\it negative} 
spin current for a positive external field $B> 0$. The negative sign is 
attributed to the dominant, negative weight of ${\rm Im}X^{-+}(\omega,k)$ 
in the positive-$\omega$ region, and (as we will show in the next section) 
the sign is opposite to that of the spin current in 3D ordered ferromagnets. 
This agrees with the experimental result of the main text. 
By using the curved spinon dispersion, we can also show that the spinon 
spin current is proportional to the external field $B$ 
in the low-field region ($|B|<J$), 
as shown in Fig. 4 of the main text. 
In Fig. 4, keeping the LSSE of Sr$_2$CuO$_3$ in mind, we set $J=-2000$ K, 
$\Delta=1$~[19, 20, 21], 
$T=20$ K and $\tau_s=1/(200$ K)~[28]. 
We emphasize that it is essential to take into account effects of 
the curved dispersion (i.e., breaking of "particle-hole" symmetry) 
in the calculation of the spin-chain spin current.

\subsection*{SC. Spin Seebeck effect of ferromagnets}
As a comparison to the spin chain, we review the theory of LSSE of 
three-dimensional (3D) ordered ferromagnetic insulators~[28]. 
Let us consider a simple Heisenberg ferromagnet on cubic lattice. 
Its Hamiltonian is given by
\begin{eqnarray}
{\cal H}_{\rm 3D}&=& 
-J \sum_{\langle \bol r,\bol r'\rangle} 
{\bol S}_{\bol r}\cdot{\bol S}_{\bol r'}
-D_z\sum_{\bol r} (S_{\bol r}^z)^2
-B\sum_{\bol r}S^z_{\bol r},
\label{eq:3dferro}
\end{eqnarray}
where ${\bol S}_{\bol r}$ is the spin-$S$ operator on site $\bol r$, $J>0$ 
is the ferromagnetic exchange coupling constant, $D_z>0$ is the easy-axis 
anisotropy. If we replace $N^{-1}\sum_k X^{-+}(\omega,k)$ with the 
susceptibility of the ferromagnet $(N_xN_yN_z)^{-1}\sum_{\bol k}
X^{-+}(\omega,\bol k)$ in Eqs.~(\ref{eq:spincurrent}) and 
(\ref{eq:susceptibility}) ($N_\alpha$ is the total number of sites 
along $\alpha$ direction), Eq.~(\ref{eq:spincurrent}) 
can be used as the formula of the spin current for the ferromagnets.

The spin-wave theory~\cite{Mattis,Yoshida,Kittel} is useful to compute the 
susceptibility of ordered ferromagnets in the low-temperature regime ($T<J$). 
According to the spin-wave theory, the Hamiltonian of Eq.~(\ref{eq:3dferro}) 
is approximated by 
\begin{eqnarray}
{\cal H}_{\rm 3D}&=& \sum_{\bol k}
\omega_{\rm sw}(\bol k)a_{\bol k}^\dag a_{\bol k},
\label{eq:spin-wave}
\end{eqnarray}
where $a_{\bol k}^\dag$ ($a_{\bol k}$) is the creation (annihilation) 
operator of a magnon with the wave vector $\bol k$, and 
$\omega_{\rm sw}(\bol k)=-2SJ(\cos k_x+\cos k_y+\cos k_z-3)+2SD_z+B$ 
is the magnon dispersion. Here we have assumed that spins are polarized along 
the direction of $B>0$. Based on the magnon (spin-wave) picture, 
the transverse spin susceptibility is viewed as the Green's function of 
magnon, and the result is written as  
\begin{eqnarray}
X^{-+}(\omega,\bol k) &=& -\frac{2S}{\omega+\omega_{\rm sw}(\bol k)+i\eta}
\label{eq:spin-wave_susceptibility}
\end{eqnarray}
with $\eta\rightarrow +0$. The imaginary part is hence given by 
${\rm Im}X^{-+}(\omega,\bol k)=2\pi S\delta(\omega+\omega_{\rm sw}(\bol k))$. 
The temperature effect in the susceptibility can be taken 
if we replace $S$ with 
$\tilde S=S-\langle a_{\bol r}^\dag a_{\bol r}\rangle$ in 
Eq.~(\ref{eq:spin-wave_susceptibility}). This equation shows that 	
the weight of ${\rm Im}X^{-+}$ is located at the negative-$\omega$ region, 
which is in sharp contrast with the result of spin chains. 
Therefore, the spin current of 3D ferromagnets has the {\it positive} sign 
which is opposite to that of spin chains. Using these results, we arrive 
at the following spin-current formula of 3D ferromagnets: 
\begin{eqnarray}
\tilde I_s = \frac{\tilde S}{T^2}
\,\,(2\pi)^{-3}\int_{-\pi}^{\pi}dk_xdk_ydk_z\,\,
\frac{\omega_{\rm sw}(\bol k)^2}{1+\tau_s^2\omega_{\rm sw}(\bol k)^2}
\frac{1}{\sinh^2(2\omega_{\rm sw}(\bol k)/T)}. 
\label{eq:spin-wave_spincurrent}
\end{eqnarray}
In a sufficiently low-temperature region, the magnon dispersion can 
be approximated by a spherical one $\omega_{\rm sw}(\bol k)\approx SJ 
|\bol k|^2 +\epsilon_0$ with spin gap $\epsilon_0=2SD_z+B$. 
In this case, the multiple integration $(2\pi)^{-3}\int dk_xdk_ydk_z$ is 
simplified to $(2\pi^2)^{-1}\int_0^{\Lambda'} dk \,\,k^2$ where $\Lambda'$ is 
the high-energy cut off of magnon band. The result is shown in 
Fig. 4 of the main text, in which we set $S=2$, $J=50$ K, $D_z=0.01 J$, 
$T=20$ K and $\tau_s=1/(200$ K)~[28]. 
The magnetic-field dependence of the calculated spin current 
is consistent with the experimental results of ferromagnets 
such as $\mathrm{Y_3Fe_5O_{12}}$~[11, 12, 54].

\end{document}